\documentclass[%
preprint,
superscriptaddress,
 amsmath,amssymb,
 aps,
 prb,
]{revtex4-2}

\usepackage{multirow}

\usepackage{graphicx}
\usepackage{dcolumn}
\usepackage{bm}

\newcommand{\MBT}{MnBi$_2$Te$_4$}

\newcommand{\PBT}{PbBi$_2$Te$_4$}

\newcommand{\XMBT}{(Mn$_{1-x}$A$^{\rm IV}_x$)Bi$_2$Te$_4$}

\newcommand{\PMBT}{(Mn$_{1-x}$Pb$_x$)Bi$_2$Te$_4$}
\newcommand{\SMBT}{(Mn$_{1-x}$Sn$_x$)Bi$_2$Te$_4$}
\newcommand{\GMBT}{(Mn$_{1-x}$Ge$_x$)Bi$_2$Te$_4$}
\newcommand{\Xelem}{A$^{\rm IV}$ = Ge, Sn, Pb}

\newcommand{\Gbar}{$\bar{\Gamma}$}

\begin{document}

\title{The electronic structure of \PMBT: experimental evidence of topological phase transition }

\author{D.\,A.~Estyunin}
\email{Contact author: estyunin@gmail.com}
\affiliation{Saint Petersburg State University, 198504 Saint Petersburg, Russia}
\affiliation{Moscow Institute of Physics and Technology, Dolgoprudny, 141700 Moscow region, Russia}

\author{T.\,P.~Estyunina}
\affiliation{Saint Petersburg State University, 198504 Saint Petersburg, Russia}

\author{I.\,I.~Klimovskikh}
\affiliation{Donostia International Physics Center, 20018 Donostia-San Sebastian, Spain}

\author{K.\,A.~Bokai}
\affiliation{Saint Petersburg State University, 198504 Saint Petersburg, Russia}
\affiliation{Moscow Institute of Physics and Technology, Dolgoprudny, 141700 Moscow region, Russia}

\author{V.\,A.~Golyashov}
\affiliation{Saint Petersburg State University, 198504 Saint Petersburg, Russia}
\affiliation{Rzhanov Institute of Semiconductor Physics, Siberian Branch, Russian Academy of Sciences, Novosibirsk 630090, Russia}

\author{K.\,A.~Kokh}
\affiliation{Saint Petersburg State University, 198504 Saint Petersburg, Russia}
\affiliation{Sobolev Institute of Geology and Mineralogy, Siberian Branch, Russian Academy of Sciences, Novosibirsk 630090, Russia}

\author{O.\,E.~Tereshchenko}
\affiliation{Saint Petersburg State University, 198504 Saint Petersburg, Russia}
\affiliation{Rzhanov Institute of Semiconductor Physics, Siberian Branch, Russian Academy of Sciences, Novosibirsk 630090, Russia}

\author{S.~Ideta}
\affiliation{Research Institute for Synchrotron Radiation Science (HiSOR), Hiroshima University, Hiroshima 739-0046, Japan}

\author{Y.~Miyai}
\affiliation{Research Institute for Synchrotron Radiation Science (HiSOR), Hiroshima University, Hiroshima 739-0046, Japan}

\author{Y.~Kumar}
\affiliation{Research Institute for Synchrotron Radiation Science (HiSOR), Hiroshima University, Hiroshima 739-0046, Japan}
\affiliation{Graduate School of Advanced Science and Engineering, Hiroshima University, Higashi-Hiroshima 739-8526, Japan}

\author{T.~Iwata}
\affiliation{Graduate School of Advanced Science and Engineering, Hiroshima University, Higashi-Hiroshima 739-8526, Japan}
\affiliation{International Institute for Sustainability with Knotted Chiral Meta Matter (WPI-SKCM$^2$), Hiroshima University, Higashi-Hiroshima 739-8526, Japan}

\author{T.~Kosa}
\affiliation{Graduate School of Advanced Science and Engineering, Hiroshima University, Higashi-Hiroshima 739-8526, Japan}

\author{T.~Okuda}
\affiliation{Research Institute for Synchrotron Radiation Science (HiSOR), Hiroshima University, Hiroshima 739-0046, Japan}
\affiliation{International Institute for Sustainability with Knotted Chiral Meta Matter (WPI-SKCM$^2$), Hiroshima University, Higashi-Hiroshima 739-8526, Japan}
\affiliation{Research Institute for Semiconductor Engineering, Hiroshima University (RISE), Higashi-Hiroshima 739-8527, Japan}

\author{K.~Miyamoto}
\affiliation{Graduate School of Advanced Science and Engineering, Hiroshima University, Higashi-Hiroshima 739-8526, Japan}

\author{K.~Kuroda}
\affiliation{Graduate School of Advanced Science and Engineering, Hiroshima University, Higashi-Hiroshima 739-8526, Japan}
\affiliation{International Institute for Sustainability with Knotted Chiral Meta Matter (WPI-SKCM$^2$), Hiroshima University, Higashi-Hiroshima 739-8526, Japan}
\affiliation{Research Institute for Semiconductor Engineering, Hiroshima University (RISE), Higashi-Hiroshima 739-8527, Japan}

\author{K.~Shimada}
\affiliation{Research Institute for Synchrotron Radiation Science (HiSOR), Hiroshima University, Hiroshima 739-0046, Japan}
\affiliation{International Institute for Sustainability with Knotted Chiral Meta Matter (WPI-SKCM$^2$), Hiroshima University, Higashi-Hiroshima 739-8526, Japan}
\affiliation{Research Institute for Semiconductor Engineering, Hiroshima University (RISE), Higashi-Hiroshima 739-8527, Japan}

\author{A.\,M.~Shikin}
\affiliation{Saint Petersburg State University, 198504 Saint Petersburg, Russia}

\date{\today}

\begin{abstract}

The investigation of methods to control and optimize the physical properties of the intrinsic magnetic topological insulator (TI) \MBT~ is a critical challenge for the development of functional materials for quantum technologies and spintronics. The promising approach is to substitute Mn atoms with Ge, Sn or Pb atoms in the \XMBT~ (\Xelem) solid solutions. This substitution enables manageable tuning of the system’s magnetic and electronic properties. In this study, we present a detailed investigation of the electronic structure evolution in \PMBT~ as a function of Pb concentration, using a variety of angle resolved photoemission spectroscopy based techniques, including photon-energy-dependent studies, spin resolved and circular dichroism (CD) measurements. Special emphasis was placed on identifying experimental evidence of the theoretically predicted topological phase transitions (TPTs) in \PMBT~ near $x\approx50$~\%. The criteria for detecting TPT include the presence or absence of topological surface states (TSS) in the electronic structure, which can be identified by their characteristic helical spin structure in spin-resolved or CD spectra, as well as the closure of the bulk band gap. Our results show that the bulk gap in the \MBT-like electronic structure decreases gradually with increasing Pb concentration up to 40~\%, where it almost closes. From 40~\% to 60~\%, the band gap remains unchanged, and above 80~\%, the \PBT-like electronic structure emerges, with the bulk gap reopening. Additionally, the TSS were detected in \PMBT~ samples with Pb concentrations up to at least 30~\% and beyond 80~\%, correlating with the regions where the bulk gap is open. However, no TSS were observed at $x\approx55$~\%, indicating that the system is in topologically distinct phase compared to \MBT~ or \PBT. At this concentration, the system may be in a semi-metallic state or a trivial insulator phase with a very narrow bulk gap. The demonstrated tunability of the electronic structure in \PMBT~ highlights its potential for further exploration in topological and spintronic applications.

\end{abstract}


\maketitle


\section{Introduction}

The potential for controlling the physical properties of the magnetic topological insulator (TI) \MBT~ is currently actively investigated \cite{Wang2021}. This TI belongs to the family of layered van der Waals materials with the spatial symmetry group R$\bar{3}$m and the general stoichiometric formula MePn$_2$Ch$_4$ (Me = Ge, Sn, Pb, Mn; Pn = Sb, Bi; Ch = Se, Te) \cite{Okamoto2012, Neupane2012, Souma2012, Kuroda2012, Shvets2017, Eremeev.jac2017, Otrokov2019, Li2019, Hao2019, Fragkos2021, Eremeev2023_Sn, McGuire2023}. In addition to the quantum anomalous Hall and axion phases predicted and observed in \MBT~ \cite{Otrokov2019, Otrokov.prl2019, Zhang2019, Gong2019, Chen2019, Deng2020, Liu2021_NatCom}, it is feasible to achieve the Dirac and Weyl semi-metallic states \cite{Lee2021}, as well as topological superconductivity \cite{McGuire2023}, in other members of MePn$_2$Ch$_4$ series.  The substitution of structural elements, including Mn, Bi, and Te, with analogous elements has demonstrated the potential for fine-tuning the properties of mixed crystals \XMBT~ (Mn$\rightarrow$\Xelem) or Mn(Bi$_x$Sb$_{1-x}$)$_2$Te$_4$ (Bi$\rightarrow$Sb) or MnBi$_2$(Te$_x$Se$_{1-x}$)$_4$ (Te$\rightarrow$Se) \cite{Wang2022}. 

It is of particular significance to consider the substitution of Mn by non-magnetic elements  \Xelem. Theoretical calculations indicate that such a substitution will result in at least one topological phase transition (TPT). This is due to the fact that although \MBT~ and A$^{\rm IV}$Bi$_2$Te$_4$ are both strong TIs, their full topological $\mathbb{Z}_2$ invariants are different: (1;000) for \MBT~ and (1;111) for A$^{\rm IV}$Bi$_2$Te$_4$ \cite{Kuroda2012, Frolov2023, Eremeev2023_Sn}. In the case of \MBT, inversion of the states of the bulk valence band (BVB) and bulk conduction band (BCB) is observed at the $\Gamma$-point of the Brillouin zone (BZ), while in the case of A$^{\rm IV}$Bi$_2$Te$_4$ compounds, this inversion occurs at the Z-point. 

Generally it is known from the literature that the changes in physical properties turn out to be almost the same for mixed crystals in which Mn is replaced by A$^{\rm IV}$ \cite{Estyunin2023, Qian2022, Changdar2023, Estyunina2023, Frolov2023, Tarasov2023, Shikin2024_arxiv}. It is therefore possible to consider these three systems together. According to density functional theory (DFT) calculations, as presented by T.~Qian~{\it et al.} \cite{Qian2022}, for \PMBT~ (Mn$\rightarrow$Pb) a state with zero bulk band gap at $x\approx44$\% and 66\% is expected. However, the electronic structure between these concentrations has not been investigated and, therefore, a detailed DFT and angle-resolved photoemission spectroscopy (ARPES) study is required to fully elucidate the effect of Mn substitution. Further, for \GMBT~ (Mn$\rightarrow$Ge) \cite{Frolov2023}, calculations carried out within the Korringa-Kohn-Rostoker (KKR) approach, which enables the calculation of a solid solution with an arbitrary distribution and degree of atom substitution, have also proposed the closure of the bulk gap at the $\Gamma$-point at $x\approx40$\%. A.~Frolov~{\it et~al.} \cite{Frolov2023} have observed that as the Ge concentration increases, the BVB and BCB edges merge along the entire path from the $\Gamma$-point to the Z-point, with the opening of the bulk band gap at $x>60$\% and parity inversion at the Z-point. The authors assumed the potential existence of a semi-metallic or trivial insulator state within the concentration range of 40 to 60\%, which was also explored in \cite{Shikin2024_arxiv}. Considering \SMBT~ (Mn$\rightarrow$Sn) a similar change in the electronic structure with merging of the edges of BVB and BCB at the $\Gamma$ and Z-points at certain concentrations of substitution was shown \cite{Tarasov2023}. Apart from that, the Weyl semi-metal state can be expected in \XMBT, but only in the case of forced ferromagnetic order in the system (the ground state has antiferromagnetic order \cite{Estyunin2023}) at almost all substitution concentrations \cite{Shikin2024_arxiv}.  

Accordingly,  a number of potential scenarios for electronic structure modifications under the increase of the  substituent atom concentration  ({\it x}) can be assumed:
\begin{enumerate}
\item At a certain concentration {\it x}, the inversion of the CB and VB states at the $\Gamma$-point vanishes, but it does not yet appear at the Z-point. Here, the system undergoes two TPTs through the trivial insulator phase $\mathbb{Z}_2=(0;000)$.
\item  The inversion of states simultaneously exists both at the $\Gamma$-point and at the Z-point within some range of concentrations {\it x}. This scenario implies two TPTs through the phase of weak TI with $\mathbb{Z}_2$ = (0;111). 
\item The inversion of the CB and VB states vanishes at the $\Gamma$ point and simultaneously appears at the Z-point. Thus only one TPT is expected and the transition should be observed in a narrow range of concentrations.   
\end{enumerate}

Currently, there are already a large number of papers in the literature where the TPT in \XMBT~ materials is predicted and analyzed by means of theoretical calculations. In this paper, we focus on the detailed analysis of the electronic structure changes of \PMBT~ at different {\it x} from experimental side. Particular attention is paid to experimental evidence of the TPT in \PMBT~ near the Pb concentrations of $x\approx50$~\%. The presence or absence of the topological surface states (TSS) in the electronic structure of the material was employed as an experimentally measurable indicator of the topological or trivial phases respectively. In order to identify the TSS, we will analyzed circular dichroism spectra and spin-resolved spectra, in which the TSS exhibits a distinctive  pattern. Moreover, the TPT criterion is the closure of the bulk band gap.  Analysis of the gap for \PMBT~ samples at different {\it x} was carried out on the basis of spectra measured at different photon energies. These allowed us to consistently estimate the bulk band gap across the entire Brillouin zone (BZ).

\section{Results and Discussion}

\begin{figure*}
\centering
\includegraphics[width=0.7\textwidth]{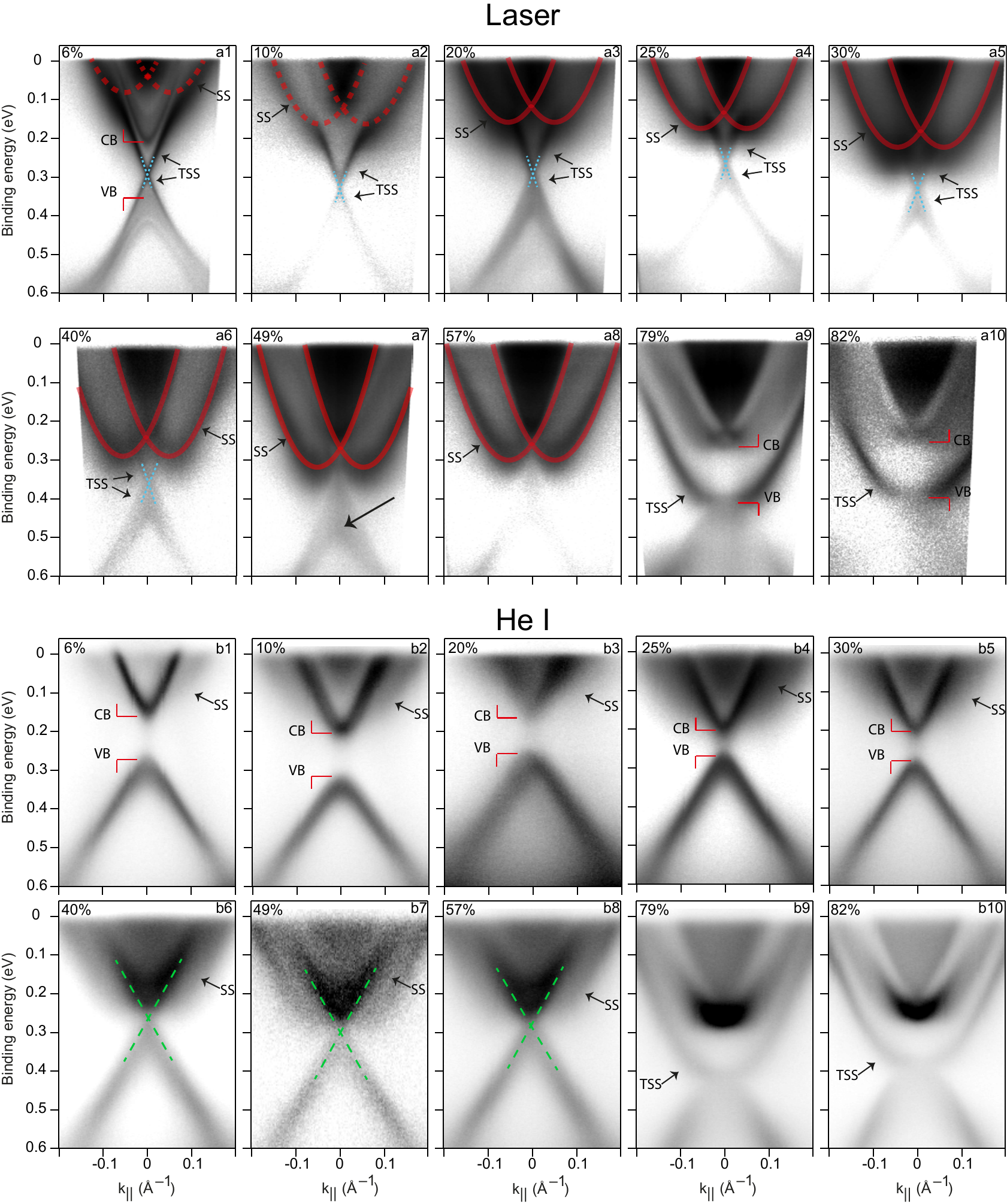}
\caption{The dispersion relations were measured under photoexcitation by light with energies of 6.3 eV [Laser] (a1 -- a10) and 21.2~eV [He I] (b1 -- b10) for \PMBT~ samples with varying {\it  x}-values. The Rashba-like states (SS) are indicated in panels a1 -- a8. Where feasible, the VB and CB edges and the TSS are indicated. The Pb molar fractions [Pb/(Pb+Mn)] shown in top left corner of the panels were estimated by the X-ray photoelectron spectroscopy and/or energy-dispersive X-ray spectroscopy methods.} 
 \label{ARPES}
\end{figure*}

\begin{figure*}
\centering
\includegraphics[width=0.7\textwidth]{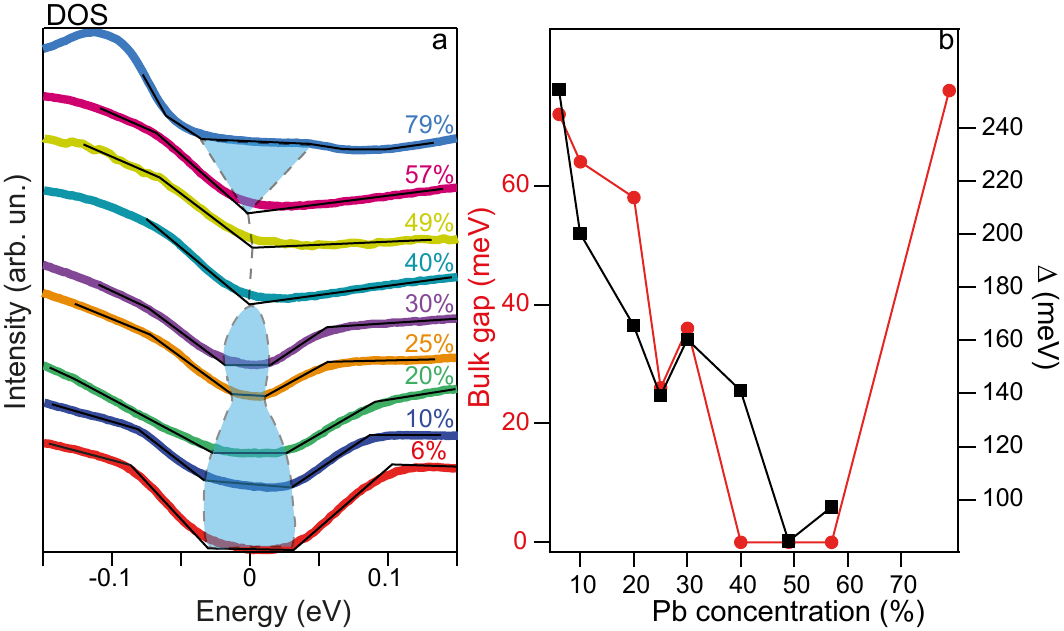}
\caption{Panel (a) illustrates the dependence of the density of states on energy [DOS(E)] for the spectra obtained at the energy of 21.2 eV. The region of the bulk band gap is indicated schematically by the dashed lines. Panel (b) illustrates the size of the bulk gap (red circles) and the difference between the center of the bulk gap and the Kramers point of the Rashba-like states [denoted as $\Delta$ = E$_{\rm G}$ - E$_{\rm KP}$] (black squares) depending on the concentration {\it x}.} 
 \label{ARPES_par}
\end{figure*}

In order to experimentally elucidate the characteristics of TPT we studied the modification in the electronic structure of \PMBT~ with increasing Pb concentration. The results of the measurements obtained by ARPES under photoexcitation by laser radiation ($h\nu$ = 6.3~eV) are presented in Fig.~\ref{ARPES}(a1 - a10), while those obtained by helium lamp radiation ($h\nu$ = 21.2~eV) are shown in Fig.~\ref{ARPES}(b1 - b10). At the energy of 6.3 eV, the TSS are clearly distinguishable in the spectra, while at 21.2 eV they are essentially undetectable, and only the BVB and BCB states are clearly visible. This effect was noted previously and related to the different photoionisation cross-section for bulk and topological states  \cite{Chen2019}. 

For the sample with Pb concentration of 6~\% in Fig.~\ref{ARPES}(a1) one can observe the CB states (at $\sim0.15-0.2$~eV), which are split due to exchange interaction (measured at T = 12~K, T$_{\rm N} \approx 24$~K) as in pure \MBT~ \cite{Estyunin2020}. Additionally, there are the VB states (at $\sim0.35-0.45$~eV), which also appear to be split. Furthermore, there are Rashba-like states (labeled as SS) near the Fermi level (schematically marked with red dashed parabolas), which are overlapped with the other CB states and the TSS \cite{Liang2022_nano, Xu2022}. Finally, there is the Dirac cone of the TSS (marked by crossing dashed lines in Fig.~\ref{ARPES}(a1 -- a6)). As can be seen, a relatively small concentration of Pb practically does not affect the \MBT~ band structure \cite{Otrokov2019}. Nevertheless, at a concentration of approximately 10~\%~(a2) -- 20~\%~(a3), modifications of the electronic structure become apparent. The Rashba-like states exhibit a shift towards higher binding energies and become clearly visible in the ARPES spectra. The TSS remain visible in the spectra, although their width slightly increases. At 30~\%~(a5) and 40~\%~(a6) the TSS are rather weakly resolvable in the spectra, and they appear blurred and overlapped by other states. At $x=49$~\%~(a7), a transformation occurs in the upper part of the VB. Although the cone-like dispersion relation is maintained, there is no distinguishable narrow bands of the TSS, while a notable density of states can be observed inside this ``cone'' (marked by the arrow in Fig.~\ref{ARPES}(a7)). At 57~\%~(a8), the intensity of cone-like VB states practically disappears, and only weak traces of these states can be observed. At these Pb concentrations the most intense states of the CB are Rashba-like states. Upon further increase of Pb concentration up to 79~\%~(a9), the electronic structure is modified dramatically and already appears similar to the electronic structure of \PBT~ \cite{Kuroda2012}. This transition from one electronic structure to another is likely to occur within a relatively narrow concentration range being discontinuous rather than a smooth transition.

The Rashba-like states appears to be an intrinsic feature of the spectra of all \PMBT~ crystals. To analyze these states, we used data from spectra in which they are clearly distinct Fig.~\ref{ARPES}(a3 -- a8). The positions of the states were approximated by model function (${\rm E_R(k)}=\frac{\hbar^2}{2m^*}k^2\pm \alpha_{\rm R} |k|$). The obtained values of the Rashba parameter ($\alpha_{\rm R} \approx 1.6$~eV$\cdot$\AA ) and the effective mass ($m^*$) coincided within the error (parabolas in panels a3 -- a8 are plotted using the averaged parameters). This can suggest that the Rashba-like states have the same orbital composition for crystals with different Mn substitution. Furthermore, the significant value of $\alpha_{\rm R}$ may indicate that these Rashba-like states originate from orbitals of Bi atoms possessing a large internal spin-orbit interaction \cite{Ishizaka2011, Bihlmayer2022}. Indeed, in the DFT calculations in \cite{Estyunina2023} one can find that Rashba-like states are mainly consisted of Bi $p_z$ density of states. Furthermore, such Rashba-like states were also found in \GMBT~ and \SMBT~ \cite{Frolov2023, Tarasov2023, Estyunina2023, Shikin2024_arxiv}. The observation of these states irrespective of the Mn substituents (Ge, Sn or Pb), for which the strength of the spin-orbit interaction differs significantly, may indicate that Rashba-like states (i.e. their parameters $\alpha_{\rm R}$ and $m^*$)  are not related significantly to the substituent atoms.

It is challenging to precisely estimate the size of the bulk gap based on the data presented in Fig.~\ref{ARPES}(a1 -- a10), primarily due to the ambiguity in the determination of the VB and CB edges. Therefore, we will analyze the dependence of the bulk gap on the Pb concentration on the basis of spectra obtained under photoexcitation by helium lamp radiation $h\nu = 21.2$~eV [Fig.~\ref{ARPES}(b1 -- b10)], as well as by synchrotron radiation (Fig.~\ref{hv_dep}). At $h\nu = 21.2$~eV, the spectrum of samples with a small Pb concentration of 6~\% to 30~\% (b1 -- b5) is dominated by bulk-like states. There are one VB and one CB state with a shape close to a cone, although with a finite value of curvature (effective mass) at the \Gbar-point. It can be seen that as the Pb concentration increases, the distance between these states decreases, resulting in a reduction in the bulk gap.  The presence of Rashba-like states is noticeable in the spectra even at low concentration, while at concentrations exceeding 25~\% (Fig.~\ref{ARPES}(b4)) they become pronounced. At concentration of $x=40$~\%~(b6), the edges of the VB and the CB come closer, forming a cone-like structure. This may indicate that the material is approaching a semi-metallic or so-called ``3D~graphene" state, as in the topological material GeSb$_2$Te$_4$ \cite{Nurmamat2020}. It can be observed that as the Pb concentration is increased from 40~\% to at least 57~\%~(b6 -- b8), the electronic structure practically does not change. At concentrations of Pb exceeding 79~\%, the dispersion relations of \PMBT~ resemble  those of \PBT~ \cite{Kuroda2012}. It is noteworthy that the dispersion relations of the bulk and surface states obtained using laser radiation Fig.~\ref{ARPES}(a9 -- a10) and helium lamp  radiation (He~I) Fig.~\ref{ARPES}(b9 -- b10) are almost identical. Thus, there is no significant dependence of the dispersion relations on the photon energy, which differs from the behavior observed in the case of the \MBT-like electronic structure [see Fig.~\ref{ARPES}(a1 -- a8) and Fig.~\ref{ARPES}(b1 -- b8)]. 

The estimation of the size of the bulk band gap was based on an analysis of the energy distribution curves obtained by integration over the momentum of the ARPES spectra in Fig.~\ref{ARPES}(b1 -- b9) and therefore to some extent similar to the density of states spectra [DOS(E)]. The curves are shown in Fig.~\ref{ARPES_par}(a). The ARPES spectra in Fig.~\ref{ARPES}(b1 -- b9) were selected for analysis as they clearly demonstrate edges of the CB and VB. We used DOS(E) spectra for analyzing the size of the bulk band gap since that enables the estimation of indirect band gaps with the minimum out of the \Gbar-point (such as in the \PBT-like electronic structure (see Fig.~\ref{ARPES}(a9)), opposite to the commonly used approach based on the estimation only of the energy distribution curves at the \Gbar-point. The bulk gap was defined as the energy region corresponding to the plateau of the DOS(E) between the linear approximations of the intensity of VB and CB states (see Fig.~\ref{ARPES_par}a). In panel (a), the region of the bulk gap is schematically marked by dashed curves connecting the edges of the gap in spectra for samples with varying Pb concentrations. The bulk gap values as a function of Pb concentration are presented in panel (b). 

It is noteworthy that as the Pb concentration increases, the energy position of the Rashba-like states with respect to the center of the bulk gap (approximately coinciding with the position of the DP of the TSS when possible) also rises as shown in Fig.~\ref{ARPES_par}b (black squares). However, the energy position of all other states remains almost unaffected by the concentration variation. This indicates that the electron filling of the Rashba-like states occurs independently of all other states. 

The obtained data show that as {\it x} increases, the size of the bulk band gap decreases from approximately 70~meV at $x = 6$~\% to almost 0~meV at $x = 40$~\%, then there is a plateau at least up to the concentration $x = 60$~\%, and at $x\approx80$~\% the band gap increases again. In general, the dependence of the band gap on Pb concentration is in a good agreement with the theoretical estimation presented in \cite{Qian2022}. The existence of a plateau in the dependence may suggest that scenarios (1) or (2)  are more probable, implying the occurrence of two TPTs and the existence of an additional phase within  the continuous Pb concentration range.

\subsection{Photon energy dependence}

\begin{figure*}
\includegraphics[width=0.75\textwidth]{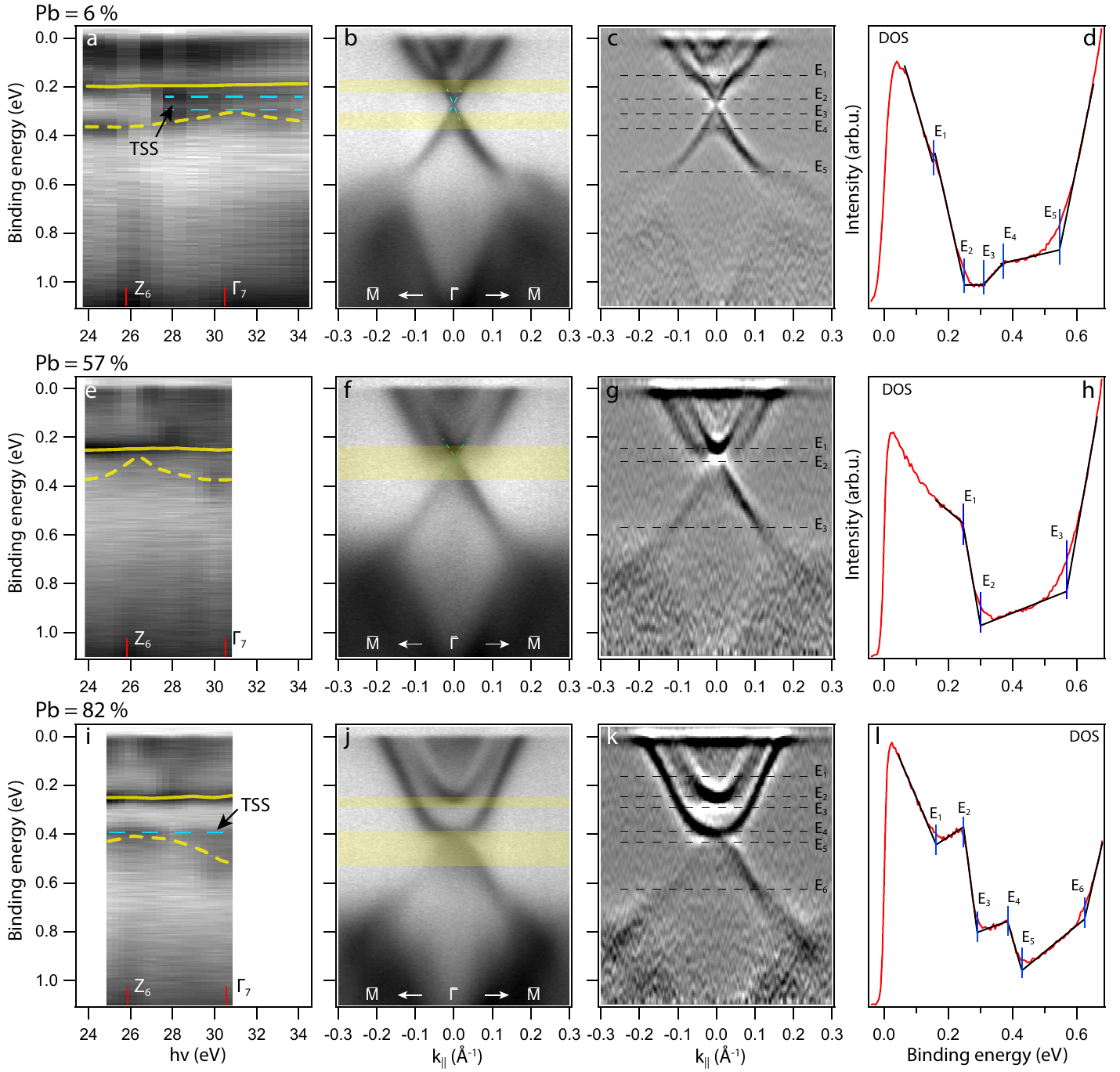}
\caption{(a,e,i) -  the set of profiles at the \Gbar-point measured at different photon energies. Yellow lines schematically mark the edges of VB (dashed) and CB (solid). In panels (a) and (i), horizontal  dashed blue lines  mark the TSS. The position of the bulk $\Gamma$ and Z points and the correspondence between the value of the wave vector and the $h\nu$ were extrapolated from the data in \cite{Hao2019}. (b, f, j) -  dispersion relations obtained by summing the spectra at different photon energies within the energy ranges represented in panels (a,e,i). Yellow stripes schematically mark the edges of the bulk band gap. In panels (c, g, k) the corresponding dispersion relations are represented in the form of second derivatives $d^2E/dk^2$. Dashed lines mark the energies at which jumps are observed in the DOS(E) spectra, presented in panels (d, h, l). All measurements were performed at T~=~20~K.}
\label{hv_dep}
\end{figure*}   

In order to analyze the bulk band gap at different points of the  BZ in the $\Gamma$-Z direction (dependence on the wave vector perpendicular to the surface, $k_z$), we considered the dependence of ARPES spectra on photon energy (Fig.~\ref{hv_dep}). The measurements were conducted on crystals with Pb concentrations of 6~\% (\MBT-like structure, Fig.~\ref{hv_dep}a-d), 57~\% (TPT region, Fig.~\ref{hv_dep}e-h), and 82~\% (\PBT-like structure, Fig.~\ref{hv_dep}i-l). For each sample, there are the photon energy dependence of the intensity  profiles at the \Gbar-point of the 2D~BZ (normal emission), the dispersion relations obtained by summing the spectra at different photon energies, their second derivatives $d^2E/dk^2$ and the corresponding DOS(E) spectra.  In this case, the DOS(E) were obtained by summing over all components of the wave vector, thereby enabling an estimation of the bulk band gap in the 3D~BZ.

For all three crystals (panels a, e, i) at the binding energy of about 0.2~eV an intense state corresponding to the CB edge (marked by the solid curve in yellow) can be identified. The state at the CB edge has a similar appearance in crystals at all three concentrations and exhibits minimal dependence on $k_z$ (or photon energy), which aligns with the findings of DFT calculations for \PBT~\cite{Kuroda2012} and \MBT~\cite{Li2019}.  This also indicates the potential existence of a state with quasi 2D character or even surface localization (surface resonance) at the CB edge, which was also observed in the results of DFT calculations in paper \cite{Shikin2024_arxiv}. 

The identification of the upper edge of the VB was a considerably more challenging process. Firstly, the edge location is dependent on $k_z$ and therefore exhibits a change in position at different values of $h\nu$. Secondly, the photoionization cross-section changes, and at specific energies the VB edge is barely distinguishable.  In order to identify the VB edge, we also relied on a number of papers that present theoretical calculations for similar structures \cite{Kuroda2012, Li2019, Frolov2023, Wang2022}. However, even within the same paper \cite{Frolov2023}, DFT calculations show a strong dependence of the VB and CB edges on $k_z$, while KKR calculations (at $x<80$~\%) show nearly flat bands in the $\Gamma$Z direction.  The resulting estimation of the position of the edge is shown schematically by the dashed curve in yellow in panels (a, e, i). However, this estimate is rather approximate.

The photon-energy-dependent ARPES spectra for samples with a Pb concentration of 6~\% and those for \MBT~ samples (see \cite{Chen2019, Estyunin2020, Frolov2023}) exhibit considerable resemblance. The minimum of bulk band gap is observed in the vicinity of the $\Gamma$-point, while the maximum is observed in the vicinity of the Z-point. Furthermore, it was determined that at photon energies exceeding 28~eV within the bulk band gap, there are states (indicated by an arrow in panel (a)) that are likely to be TSS. It can be observed that these states remain constant with respect to changes in photon energy (marked by two parallel dashed lines in blue). It is important to note that the observed gap in the TSS can be attributed to the enhanced angular acceptance during the plotting of the $h\nu$-dependent spectrum. Overall, one can conclude that TSS in \MBT-like samples can be resolved also at energies above 28~eV, which has not been shown before, as well as at energies below 16~eV which was shown in~\cite{Chen2019}. 

Panel (b) shows the spectrum obtained by summing several spectra at photon energies from 24 to 35~eV. It is evident from the spectrum (b) and its second derivative (c) that cone-like states (illustrated by dashed crossing lines in blue) are present within the bulk band gap, which is indicated schematically by horizontal stripes. The finding, presented at the DOS(E) spectrum (d), reveals a plateau region between the energies E$_2$ and E$_3$ (E$_3$ - E$_2$ = 61~meV). The size of this region gives an estimate  for the size of the bulk band gap. The resulting value is slightly smaller than that (about 72~meV) obtained previously from the spectrum at $h\nu=21.2$~eV  (see Fig.~\ref{ARPES}b1 and Fig.~\ref{ARPES_par}b). 

At the Pb concentration of {\it x} = 57~\%, the minimum between VB and CB  appears to shift to the Z-point (panel e). At this concentration, the electronic structure of the mixed crystal is rather between the two TPT points \cite{Qian2022}. In the DOS(E) spectrum (panel (h)), as well as in the case of spectra for samples with Pb concentrations 40~\% - 57~\% in Fig.~\ref{ARPES_par}a, no regions with intensity plateaus can be distinguished.  Furthermore, the spectrum in panel (f) demonstrates that the bulk states exhibit a conical dispersion relation (illustrated by dashed lines in green). This supports the hypothesis that the electronic structure is in proximity to the semi-metallic state (i.e. the bulk gap is narrow enough that cannot be distinguished). 

At Pb concentration of 82~\%, the electronic structure takes the form of a \PBT-like electronic structure (panel j). In this case, as was also observed for high Ge concentrations in \GMBT~ \cite{Frolov2023}, there are the TSS within the bulk gap region, with the DP overlapping with the edge of the VB (schematically marked by dashed horizontal line in blue). In the DOS(E) spectrum (panel l), a linearly dependent region between the energies E$_3$ and E$_4$ (E$_4$ - E$_3$ = 95~meV)  can be identified. This region is expected to correspond to the bulk band gap, in accordance with \cite{Kuroda2012}. It is noteworthy that we attribute the linear dependent region to the bulk band gap and observe no plateau region there. This is due to the fact that in the case of a \PBT-like electronic structure the TSS exhibit  strong warping and their dispersion relation is considerably distant from the conical one. While conical dispersion relation results in the intensity plateau during integration along wave vector. 

The photon energy-dependence data support the conclusion drawn from the data in Fig.~\ref{ARPES}.  It is noteworthy that for crystals with a substitution ratio of approximately 60~\%, the bulk band gap is essentially absent, and their electronic structure is close to the semi-metallic state.  Given that the dispersion relations for crystals with a Pb concentration ranging from 40~\% to 60~\% [Fig.~\ref{ARPES}(b6 -- b8)] exhibit near-identical characteristics, it can be assumed that the region of substitution concentrations in which the states close to semi-metallic state are observed is quite wide. This conclusion aligns with the DFT estimations  presented in \cite{Qian2022}. Further attempts based on the analysis of  spin-resolved spectra and circular dichroism spectra will be made to identify additional characteristics of the electronic structure of the mixed crystals and confirm the TPT. 

\subsection{Circular dichroism}

\begin{figure*}
\includegraphics[width=0.72\textwidth]{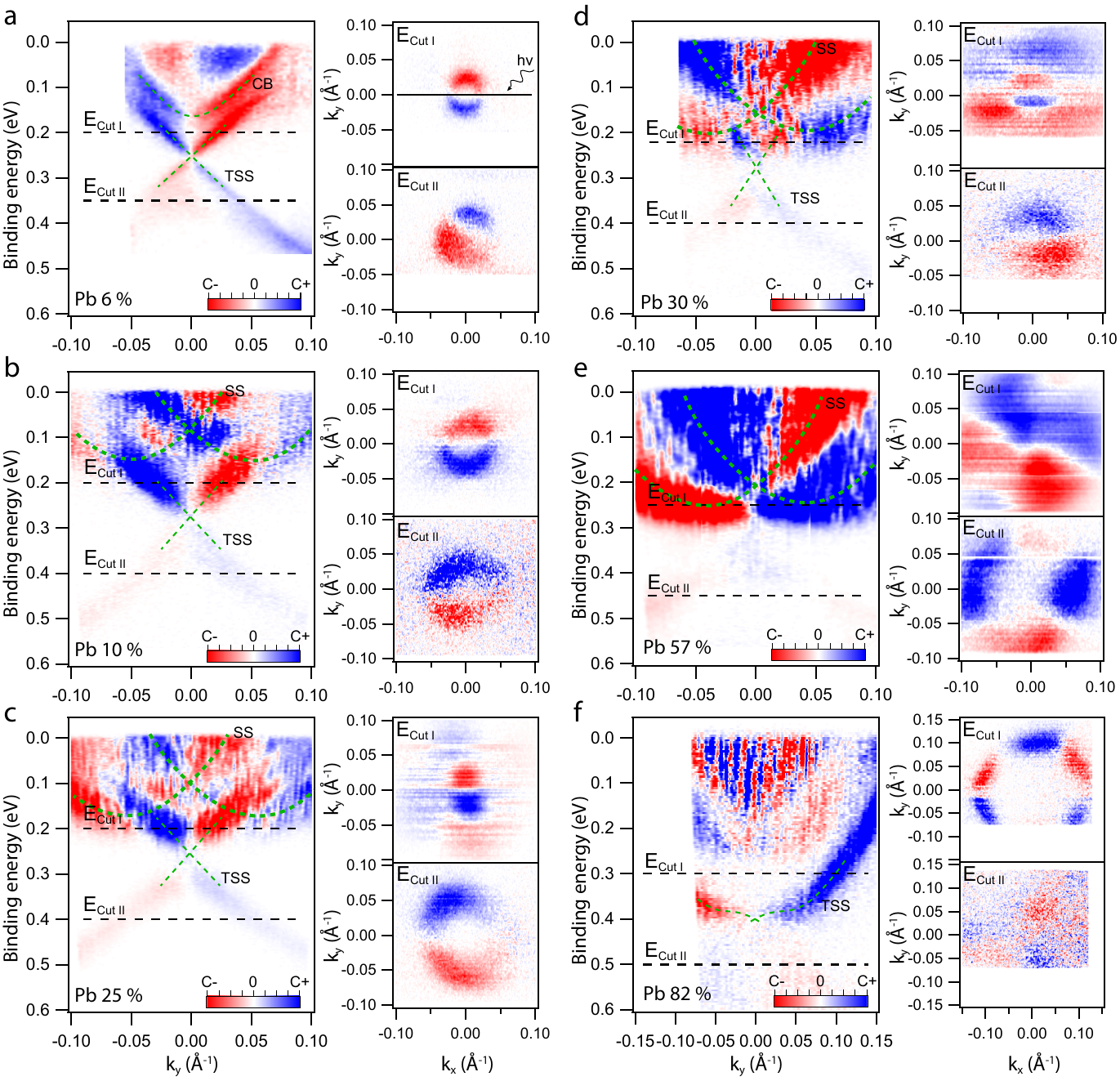}
\caption{Spectra of circular dichroism. Each panel shows the spectrum of I(E,$k_y$) on the left, while on the right side, there are slices at constant energy I($k_x$, $k_y$) taken in the CB (Cut I) and VB (Cut II) regions. The $x$-axis is aligned with the analyzer slit and the plane of light incidence, while the $y$-axis is perpendicular to these. The green dashed lines schematically indicate the position of the TSS, the Rashba-like states (SS) and the CB edge, where possible. The measurements were performed at energy of $h\nu=6.3$~eV. }
\label{dich}
\end{figure*}  

The circular dichroism (CD) method is a widely employed technique for the identification of the TSS in dispersion relations obtained by means of ARPES. The method implies the analysis of the difference between the spectra obtained through photoexcitation by light with right (c+) and left (c-) circular polarizations (${\rm I}_{\rm CD} = {\rm I}_{\rm c+} - {\rm I}_{\rm c-}$) \cite{Wang2011_CD, Wang2013_CD}. In the CD spectra, the TSS exhibits a distinctive feature: the dispersion relations I$_{\rm CD}$(E,k) for the opposing branches of the TSS cone display an inversion of the intensity sign with respect to the \Gbar-point.  The characteristic features are also observed for other states possessing spin polarization, for example, for Rashba states. While the CD ARPES is rather complicated and depends on the photoemission matrix elements and photon energy \cite{Scholz2013a} the strong CD signal of the surface states can be considered as a consequence of the spin polarization under some conditions \cite{Li2021}. 

Fig.~\ref{dich} shows the results of the study of \PMBT~ samples by means of CD ARPES. The CD spectrum of the sample with a Pb concentration of $x=6$~\% (panel a) exhibits a structure characteristic of the TSS. Furthermore, on the cuts at constant energy, a circular contour that changes sign when passing through $k_x=0$~\AA$^{-1}$~ (plane of light incidence) is clearly visible. This is also a characteristic feature of the  TSS in the CD \cite{Wang2013_CD}. This confirms the correctness of the identification of these states as the TSS previously made based on Fig.~\ref{ARPES}(a1). It is noteworthy that the CB states also exhibit a non-zero signal in the CD spectra. Usually, the bulk states exhibit no intensity in the CD spectra, and the majority of the signal is coming from the surface states as in the case of Bi$_2$Se$_3$ \cite{Wang2013_CD}. Therefore, this may indicate the presence of additional surface states at the CB edge, as was previously observed in $h\nu$-dependent spectra (see Fig.~\ref{hv_dep}) and DFT calculations \cite{Shikin2024_arxiv}. 

As the Pb concentration increases (10~\%, panel b), the cone-shaped states maintain their CD pattern corresponding to the TSS. Furthermore, the Rashba-like states appear in the spectrum  (marked by two parabolas, see dashed lines). The distinctive pattern for them observed in the CD spectra is the presence of two parabolas of different color i.e. opposite sign of intensity. Upon further increase of the Pb concentration to 25~\% (c) and 30~\% (d), a notable decrease in the CD intensity of the cone-shaped states and an accompanying increase in the CD intensity of the Rashba-like states are observed. It is noteworthy that the CD intensity is opposite for the TSS and outer branches of Rashba parabolas. For example, where I$_{\rm CD}$ for the cone is greater than zero (blue), it is negative (red) for the outer branches of Rashba-like states. 

At a Pb concentration of 57~\% (panel e), the specific pattern of the TSS in CD spectra is absent, allowing us to conclude that there is no TSS present in the electronic structure. Consequently, \PMBT~ at this concentration is no longer a strong TI with $\mathbb{Z}_2$=(1;000) and the TPT occurs at lower Pb concentrations.  Meanwhile, it is evident that the pattern of the Rashba-like states persists in the CD spectra. At a Pb concentration of 82~\% (panel f), the CD spectra again demonstrate a state with the pattern characteristic of the TSS. However, here the pattern has the triple alternation of the CD intensity sign when going around the $\Gamma$-point (see the constant energy profiles) due to the different orientation of the crystal according to \cite{Wang2011_CD}. It is also worth noting that, only the top part of the TSS cone is visible, as the Dirac point is located within the BVB.

Based on the analysis of CD spectra, we assume that the TSS can be identified in samples with Pb concentrations of up to 30~\% (phase of strong TI with $\mathbb{Z}_2$=(1;000)) and in samples with Pb concentration exceeding 80~\% (phase of strong TI with $\mathbb{Z}_2$=(1;111)). At a Pb concentration of approximately 57~\%, no states that can be identified as the TSS are observed. To further validate the findings, spin-resolved spectra for the mixed crystal with the such Pb concentration  were considered. 

\subsection{Spin resolved measurements}

\begin{figure*}
\includegraphics[width=0.75\textwidth]{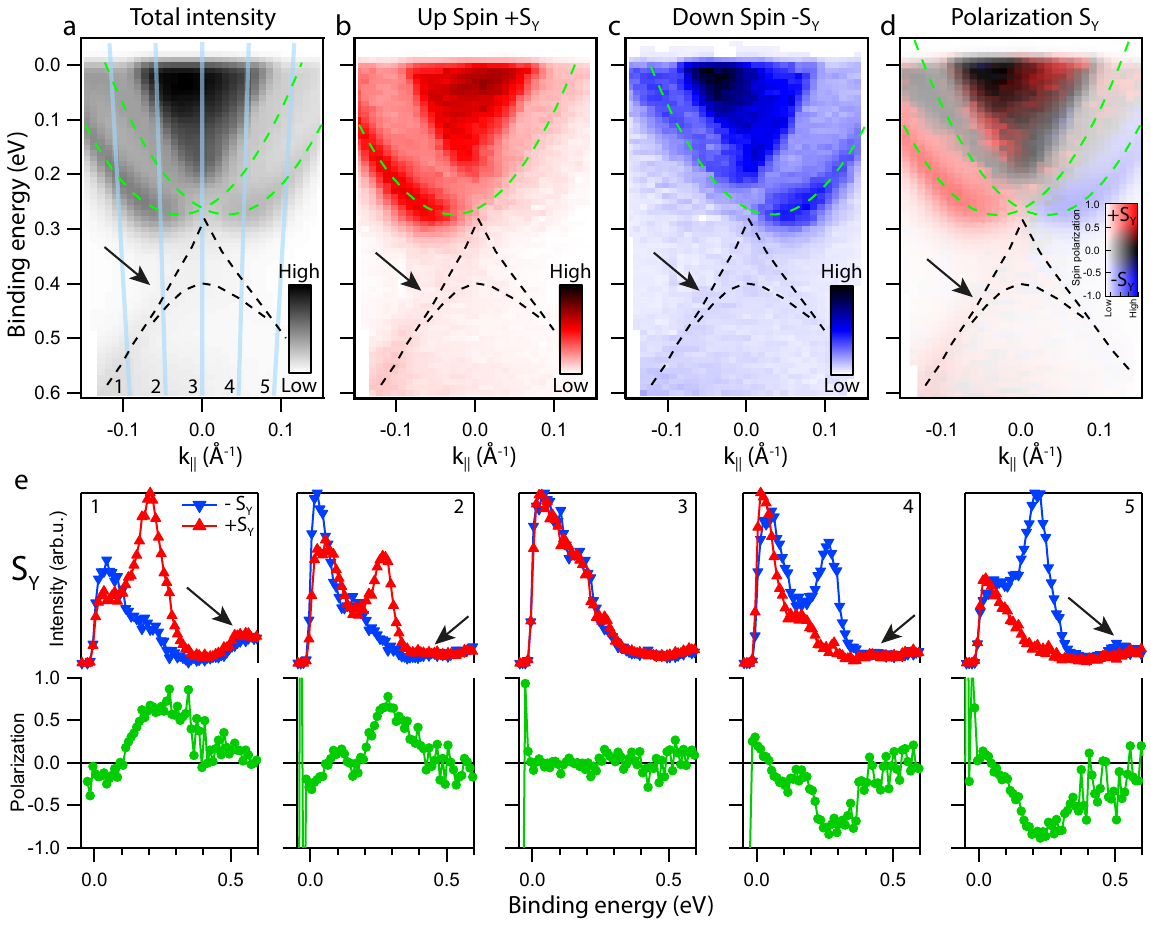}
\caption{The dispersion relations for a mixed crystal with Pb concentration of 54~\% measured with spin resolution for the S$_Y$ spin component (in-plane, perpendicular to the momentum). Panel (a) - the total intensity, panel (b) and (c) - projection on $+{\rm S}_Y$ and $-{\rm S}_Y$ spin,  panel (d) -  the spin polarization. The dashed lines schematically mark the states of interest observed in the spectra. The spin polarization spectrum is plotted with the two-dimensional color scale shown, with the horizontal axis corresponding to intensity and the vertical axis corresponding to spin polarization. (e) Spin-resolved energy profiles with projection onto the spin components $+{\rm S}_Y$ (red) and $-{\rm S}_Y$ (blue) and the corresponding spin polarization of states at several electron emission angles (shown in panel (a)).}
\label{Spin}
\end{figure*}

Spin-resolved measurements were conducted on a sample with a Pb concentration of 54~\%. Fig.~\ref{Spin} illustrates the dispersion relations obtained for the in-plane (perpendicular to the momentum) spin projections $+{\rm S}_{\rm Y}$ (b) and $-{\rm S}_{\rm Y}$ (c), as well as the spin polarization of the states (d). The principal objective was to identify the manifestation of the TSS in the VB region, where the TSS do not overlap with Rashba-like states. As also demonstrated in the measurements of samples with $x\approx50-60$~\% (see Fig.~\ref{ARPES}(a8), Fig.~\ref{hv_dep}f, Fig.~\ref{dich}e), the VB region exhibits states with a cone-shaped dispersion relation (indicated by black dashed lines and the arrow in Fig.~\ref{Spin}(a -- e)). However, it is evident from panels b, c and d that there are no states with a notable degree of spin polarization in this region.

To enhance the precision of the estimation, additional spectra with better statistics were acquired at multiple photoemission angles (panel e). The state of interest in the VB can be observed in the spectra (indicated by the arrow). Nevertheless, it can be seen that the state has a low level of spin-polarization. The polarization is around 0.1 (for panel e1, $-k_\|$) and around $-0.2$ (for panel e5, $+k_\|$).  In  comparison, the level of in-plane spin polarization for the TSS in Bi$_2$Te(Se)$_3$ is greater than 0.5 \cite{Souma2011, Kuroda2016}. Notably, certain spin polarization can appear even for the bulk-related states due to the spin-dependent interference on the surface \cite{Kimura2010, Rybkin2012}. Finally, we can conclude that the observed cone-like states in the VB are not related to the TSS. 

Additionally, the Rashba-like states were identified in spin-resolved spectra. They are schematically represented by dashed parabolic curves of green color in panels (a, b, c, d). The curves were obtained from the approximation of the experimental data points by the model function for the Rashba states. It is notable that the outer regions of the parabolas exhibit considerable spin polarization, whereas the inner sections are weakly resolved and polarized. The latter may be due to the presence of several states in this region with opposite spin values. The signals from them are mutually compensated. Fig.~\ref{Spin}e shows that the polarization of the Rashba-like states (in the region where they are well resolved and separated from other states) is approximately $0.7-0.8$, which agree with the reported in-plane spin-polarization of the Rashba states in Bi$_2$Te$_3$ \cite{Bahramy2012}.

Moreover, we would like to present some considerations regarding additional experimental approaches to studying the electronic properties of \XMBT. Most of the techniques included in the APRES method were used in this work. Further clarification of the electronic properties of \PMBT~ near the TPT ($x\approx40-60$~\%) region requires transport measurements. A number of studies have already presented transport measurements for mixed crystals with Mn substituted by  Sn or Pb \cite{Qian2022, Changdar2023}. However, the mentioned studies investigated crystals with strong n(p)-doping (in the degenerate semiconductor state), in which conduction due to trivial bulk states dominates. A potential direction for further research in this field would be the synthesis of (Mn,A$^{\rm IV}$)(Bi,Sb)$_2$Te$_4$ crystals with the substitution of Bi by Sb, with the aim of achieving the compensated semiconductor state, as for Mn(Bi,Sb)$_2$Te$_4$ crystals \cite{Chen_Nat_Com_2019, Glazkova2022}. In this case, the semi-metallic state may be distinguished from the trivial insulator state by the more precise transport characteristics. 

An additional approach for verifying the phase of the material in the vicinity of the TPT region ($x\approx40-60$~\%) is to study the electronic structure I(E,k$_x$,k$_z$) of the non-natural cleavage plane (i.e. 10$\bar{1}$1). The features expected for the Weyl semi-metal are the presence of Fermi arches along the \Gbar$\bar{\rm Z}$ direction, while a marker of a weak TI is the emergence of quasi-1D spin-polarized states along this direction, as observed for the known weak TI ZrTe$_5$ \cite{Zhang2021}. Nevertheless, this task proves to be  challenging and necessitates the creation of a measurement methodology for the non-natural cleavage plane, which has not yet been demonstrated for  Bi$_2$Te$_3$-like TIs. 

\section{Conclusion}

In conclusion, this study demonstrates an evolution of the electronic structure in \PMBT~ as the Pb concentration ({\it x}) increases from 0 to approximately 60~\%, maintaining the \MBT-like electronic characteristics. A transformation to a \PBT-like electronic structure is expected between approximately 60~\% and 80~\% Pb, which maintains at higher concentration. From 0 to around 40~\% Pb,  a smooth decrease in the size of the bulk band gap is also observed.  In the concentration range of $\sim$40~\% to $\sim$60~\%, the band gap remains constant, with the VB and CB edges approaching each other, forming a cone-like dispersion relation. The topological phase in this concentration range is ambiguous. According to the ARPES data, the system appears to be in a semi-metallic state, as no discernible bulk band gap is observed. However, theoretical predictions suggest the presence of either a trivial insulator or weak TI phase, which would imply the existence of a band gap. It is possible that the gap is smaller than the experimental resolution. With greater certainty, it was shown through CD and spin-resolved ARPES that there are no TSS in \PMBT~ at this concentration.  While the presence of the TSS was confirmed in mixed crystals with Pb concentrations below 30~\% and above 80~\%. This confirms that the material at $x\approx55$~\% exists in a distinct topological phase, and at least two TPT occur near this Pb concentration. The presented results show the broad tunability of the electronic structure of \PMBT, which provides the potential to utilize its semiconducting and topological properties in electronic applications, in addition to its well known thermoelectric properties.

\section{Experimental Section}

{\bf Sample Synthesis:} 
The bulk crystals of \PMBT~ were synthesized using a modified Bridgman method \cite{Kokh2014}. High-purity elementary Mn, Bi, Te and Pb (4N purity) were placed into conical quartz ampoules and sealed under a vacuum of approximately 10$^{-2}$ torr. The materials were then heated to 1050~$^\circ$C in a growth furnace to synthesize the charge, with compositions following the ratio \PMBT~$+$~2$\times$Bi$_2$Te$_3$. After one day of homogenization, the ampoule was moved to a cooler zone (600~$^\circ$C) at a rate of 10 mm/day.

The monocrystals obtained from the ingot were initially characterised by XPS  at ISP SB RAS (Novosibirsk, Russia) and energy-dispersive x-ray (EDX) spectroscopy  at the research resource centre of SPBU. The total molar fraction of Mn and Pb exceeded 10\%, indicating the prevalence of the \PMBT~ phase.

{\bf ARPES measurements:}  ARPES data were recorded using (i) laboratory-based facility SPECS GmbH ProvenX-ARPES system located in ISP SB RAS (Novosibirsk, Russia) equipped with ASTRAIOS 190 electron energy analyzer, a non-monochromated He I$\alpha$ light source with h$\nu$=21.2 eV and sample cooling by liquid nitrogen [Fig.~\ref{ARPES}(b1-b10)]; (ii) $\mu$-LaserARPES facility at the research institute for synchrotron radiation science (HiSOR) at Hiroshima University (Japan) equipped with Scienta R4000 analyzer, laser light source with h$\nu$=6.3 eV and sample cooling by liquid helium [Fig.~\ref{ARPES}(a1-a10), Fig.~\ref{dich}] \cite{Iwasawa2017};  (iii) syncrotron based facility BL1 end-station at HiSOR at Hiroshima University (Japan) [Fig.~\ref{hv_dep}]. Spin resolved spectra were measured  at SpinLaserARPES facility at HiSOR at Hiroshima University (Japan) equipped with Scienta DA30 analyzer, VLEED type spin detector, laser light source with h$\nu$ = 6.39~eV and sample cooling by liquid helium (Fig.~\ref{Spin}) \cite{Iwata2024}.

A clean surface was prepared by tape-cleavage inside a vacuum chamber with a pressure below 10$^{-8}$~torr. The base pressure during the measurements remained below $5\times10^{-11}$~torr.

{\bf Contribution} 

The manuscript was written and edited by D.A. Estyunin. Experimental data processing and figure preparation by D.A. Estyunin and T.P. Estyunina. ARPES measurements at ISP SB RAS were performed by D.A. Estyunin, T.P. Estyunina, A.M. Shikin, K.A. Bokai, V.A. Golyashov and O.E. Tereshchenko. ARPES measurements at HiSOR facilities (BL1, $\mu$-ARPES and SpinLaser ARPES) were performed by D.A. Estyunin, T.P. Estyunina, A.M. Shikin, I.I. Klimovskikh, S. Ideta, Y. Miyai, Y. Kumar, T. Iwata, T. Kosa, K. Kuroda, T. Okuda, K. Miyamoto and K. Shimada. Crystals provided by K.A. Kokh and O.E. Tereshchenko. Funding support by A.M. Shikin. The project was planned and supervised by D.A. Estyunin and A.M. Shikin.  All authors have read and approved the published version of the manuscript.

{\bf Funding} 

This work was supported by the Russian Science Foundation grant No. 23-12-00016 and the St. Petersburg State University grant No. 95442847.

{\bf Acknowledgements} 

ARPES measurements at HiSOR were performed under Proposals No. 23AG008, 23BU003, 23AU012, 23AU009. We are grateful to the N-BARD, Hiroshima University for liquid He supplies. The authors acknowledge the ``Center for Nanotechnology'' of the Research Park of St. Petersburg University, where the elemental composition of the samples was studied. Crystal growth was performed under state assignment of IGM SB RAS No. 122041400031-2.

\end{document}